\documentclass[conference]{IEEEtran}
\IEEEoverridecommandlockouts
\usepackage{cite}

\usepackage{amsmath,amssymb,amsfonts}
\usepackage{algorithmic}
\usepackage{graphicx}
\usepackage{textcomp}
\usepackage{xcolor}
\usepackage{tabularx}  
\usepackage{float}  
\usepackage{multirow}
\usepackage{tcolorbox}
\usepackage{booktabs}
\usepackage{soul,xcolor}
\usepackage[a-1b]{pdfx}
\usepackage{color}
\definecolor{linkcolor}{rgb}{0,0,0.7}

\sethlcolor{yellow!20}
\AtBeginDocument{
    \setlength{\abovedisplayskip}{8pt}
    \setlength{\belowdisplayskip}{6pt}
}
\begin{document}

\title{HCMA-UNet: A Hybrid CNN-Mamba UNet with Axial Self-Attention for Efficient Breast Cancer Segmentation}


\author{
  \IEEEauthorblockN{1\textsuperscript{st} Haoxuan Li\textsuperscript{*}}
  \IEEEauthorblockA{
  Shenzhen International Graduate School\\
  Tsinghua University\\
  Shenzhen, China\\
  li-hx24@mails.tsinghua.edu.cn}
  \and
  \IEEEauthorblockN{2\textsuperscript{nd}Wei Song\textsuperscript{*}}
  \IEEEauthorblockA{
  School of Automation\\
  Guangdong University of Technology\\
  Guangzhou, China\\
  sonwe@mail2.gdut.edu.cn}
\and
  \IEEEauthorblockN{3\textsuperscript{rd} Peiwu Qin\textsuperscript{†}}
  \IEEEauthorblockA{
  Shenzhen International Graduate School\\
  Tsinghua University\\
  Shenzhen, China\\
  pwqin@sz.tsinghua.edu.cn}
  \and
  \IEEEauthorblockN{4\textsuperscript{th} Xi Yuan\textsuperscript{†}}
  \IEEEauthorblockA{
  Shenzhen International Graduate School\\
  Tsinghua University\\
  Shenzhen, China\\
  yuanx20@mails.tsinghua.edu.cn}
  \and
  \IEEEauthorblockN{5\textsuperscript{th} Zhenglin Chen\textsuperscript{†}}
  \IEEEauthorblockA{
  Zhejiang Key Laboratory of Imaging and Interventional Medicine\\
  The Fifth Affiliated Hospital of Wenzhou Medical University\\
  Lishui, China\\
  chenzlin1992@163.com}
\thanks{\textsuperscript{*}These authors contributed equally to this work.}
\thanks{\textsuperscript{†}Corresponding authors.}
}
\maketitle
\begin{abstract}
Breast cancer lesion segmentation in DCE-MRI remains challenging due to heterogeneous tumor morphology and indistinct boundaries. To address these challenges, this study proposes a novel hybrid segmentation network, HCMA-UNet, for lesion segmentation of breast cancer. Our network consists of a lightweight CNN backbone and a Multi-view Axial Self-Attention Mamba (MISM) module. The MISM module integrates Visual State Space Block (VSSB) and Axial Self-Attention (ASA) mechanism, effectively reducing parameters through Asymmetric Split Channel (ASC) strategy to achieve efficient tri-directional feature extraction. Our lightweight model achieves superior performance with 2.87M parameters and 126.44 GFLOPs. A Feature-guided Region-aware loss function (FRLoss) is proposed to enhance segmentation accuracy. Extensive experiments on one private and two public DCE-MRI breast cancer datasets demonstrate that our approach achieves state-of-the-art performance while maintaining computational efficiency. FRLoss also exhibits good cross-architecture generalization capabilities. The source code is available at \url{https://github.com/Haoxuanli-Thu/HCMA-UNet}.
\end{abstract}

\begin{IEEEkeywords}
Medical Image Segmentation, DCE-MRI, Mamba, Axial Self-Attention, FRLoss
\end{IEEEkeywords}

\section{Introduction}
 Breast cancer is one of the most common cancers affecting women worldwide, with high morbidity and mortality rates, posing a major threat to women's health~\cite{britt2020key}. Accurate lesion segmentation, as a cornerstone of early diagnosis and evaluation, plays a vital role in improving patient prognosis and facilitating personalized treatment plans~\cite{milosevic2018early}. Among diverse medical imaging modalities, DCE-MRI has become a key tool for the diagnosis and evaluation of breast cancer because of its excellent soft-tissue contrast and absence of ionizing radiation. However, breast cancer lesions exhibit significant heterogeneity and unclear boundaries with the surrocunding tissues, making precise segmentation challenging. Currently, clinical practice still relies mainly on time-consuming manual segmentation, which is labor-intensive and susceptible to subjective factors. Therefore, developing robust and generalizable automatic segmentation algorithms is crucial for improving clinical efficiency and diagnostic accuracy.

 Convolutional Neural Networks particularly U-Net~\cite{ronneberger2015u} have revolutionized medical image segmentation. Through its innovative symmetric U-shaped architecture, U-Net elegantly integrates encoder downsampling and decoder upsampling to achieve coarse-to-fine feature extraction and reconstruction. However, CNNs have limitations in capturing long-range dependencies and global context, which can affect their performance in segmenting breast tumors with irregular morphologies and varying sizes.
   \begin{figure*}[!htbp]
    \centering
    \includegraphics[width=1.0\textwidth]{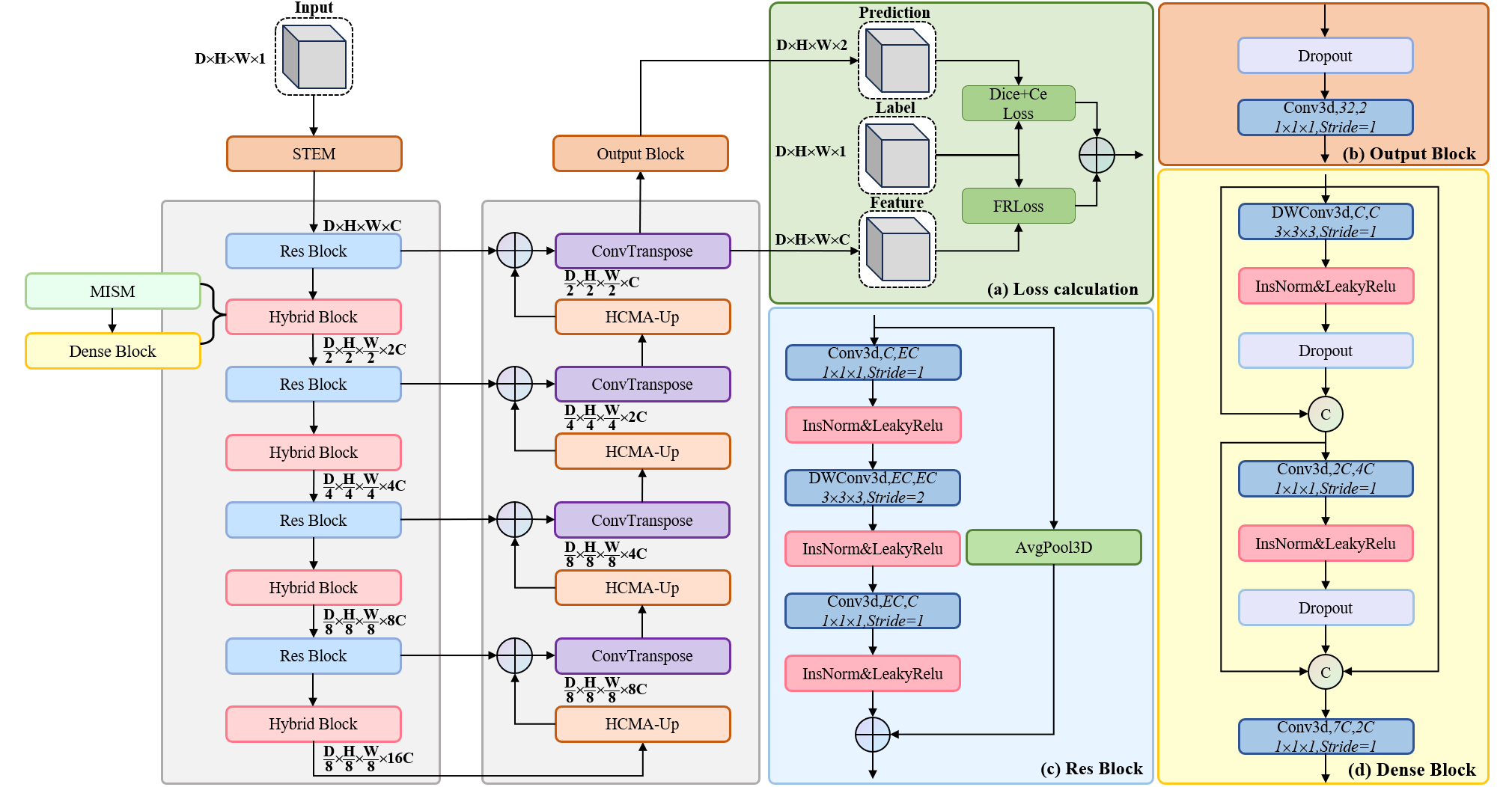}
    \caption{
    Overview of HCMA-UNet.
    (a) Loss computation module that combines Dice and Ce losses with our proposed FRLoss.
    (b) Output Block performs final classification based on the features from the last decoder layer.
    (c) Res Block for downsampling operations.
    (d) Dense Block for channel expansion.
    }
    \label{fig:HCMA}
    \vspace{-6pt}
    \end{figure*}
Transformer-based architectures have demonstrated superior performance in medical image segmentation by overcoming CNNs' receptive field limitations through self-attention mechanisms. However, transformer models inherently lack the ability to efficiently extract local features that are crucial for detailed medical image analysis. The hybrid CNN-Transformer architectures effectively address this limitation by incorporating CNNs' strong capacity for local feature extraction while maintaining transformer's global context modeling ability. Numerous successful works have demonstrated the effectiveness of this hybrid approach in medical image segmentation, such as SwinUNETR~\cite{hatamizadeh2021swin}, UNETR++~\cite{shaker2024unetr++}.
 
 Existing 3D breast cancer segmentation networks are primarily based on CNNs or Transformer architectures~\cite{zhao2024swinhr,zhou2024prototype}. However, both face trade-offs between computational efficiency and the ability to capture local and global features necessary for accurate segmentation. To overcome CNNs' local receptive field limitation of CNNs and the computational complexity of transformers, Mamba~\cite{gu2023mamba} is a novel architecture designed to achieve long sequence modeling with linear computational complexity by combining Gated Multi-Layer Perceptron (Gated MLP) with the H3 State Space Model. 

Recently, Mamba has demonstrated promising potential for 3D medical image segmentation through architectures such as U-Mamba~\cite{ma2401u}, LKM-UNet~\cite{wang2024lkm}, and MambaClinix~\cite{bian2024mambaclinix}. However, current models process 3D medical images by flattening volumetric data into sequential inputs face significant challenges, primarily because Mamba is originally designed for sequence processing tasks. When processing flattened 3D volumes as sequences, each element can only interact with previously scanned pixels through a compressed hidden state. This sequential processing mechanism imposes strict one-directional dependencies. However, such a constraint fundamentally conflicts with the non-causal nature of 3D medical data, where each voxel inherently maintains simultaneous correlations with its neighbors in all spatial directions. When such volumetric data are forcefully flattened into sequences, spatially adjacent voxels may become distant in the sequence representation, preventing the network from effectively learning their natural spatial relationships. This artificial sequential arrangement imposes an unnatural directionality on inherently omnidirectional spatial features, limiting the model's 3D structure learning capability.

To address the limitation of unidirectional modeling in 2D image segmentation, VMamba~\cite{liu2024vmamba} employs bidirectional state space models and positional embedding mechanisms. Building upon this design principle, we propose the Multi-view Axial Self-Attention Mamba (MISM) module to overcome the inherent limitations of existing 3D Mamba architectures. This module processes slices along three anatomical planes (sagittal, coronal, and axial) and models slice features using the Visual State Space Block (VSSB) from VMamba. Based on the intra-slice features extracted by VSSB, the module incorporates an Axial Self-Attention mechanism to enhance inter-slice feature correlations, achieving effective integration of both intra-slice and inter-slice information. By embedding this module into a lightweight convolutional network, we combine the local feature extraction capabilities of convolutions with the global modeling advantages of Mamba, enhancing feature learning capability while maintaining computational efficiency.
The main contributions of our work are summarized as follows:
    \begin{figure*}[!htbp]
    \centering
    \includegraphics[width=1.0\textwidth]{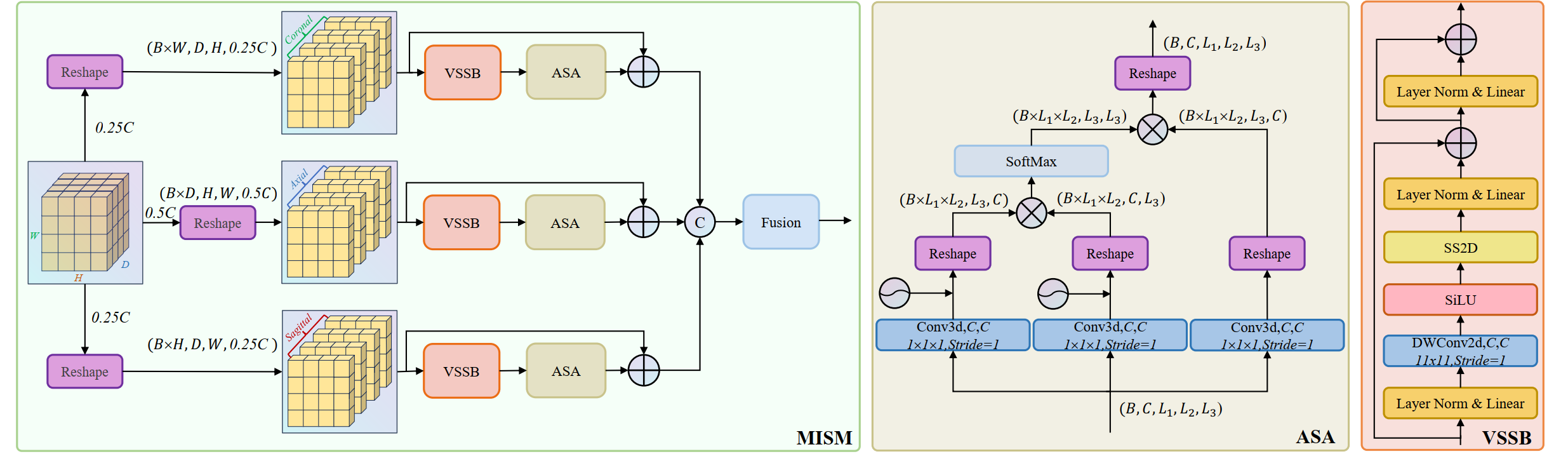}
    \caption{Overview of MISM and its sub-modules: ASA and VSSB.}
    \label{fig:MISM}
    \vspace{-5pt}
    \end{figure*}
    \begin{figure}[!htbp]
    \centering
    \includegraphics[width=0.5\textwidth]{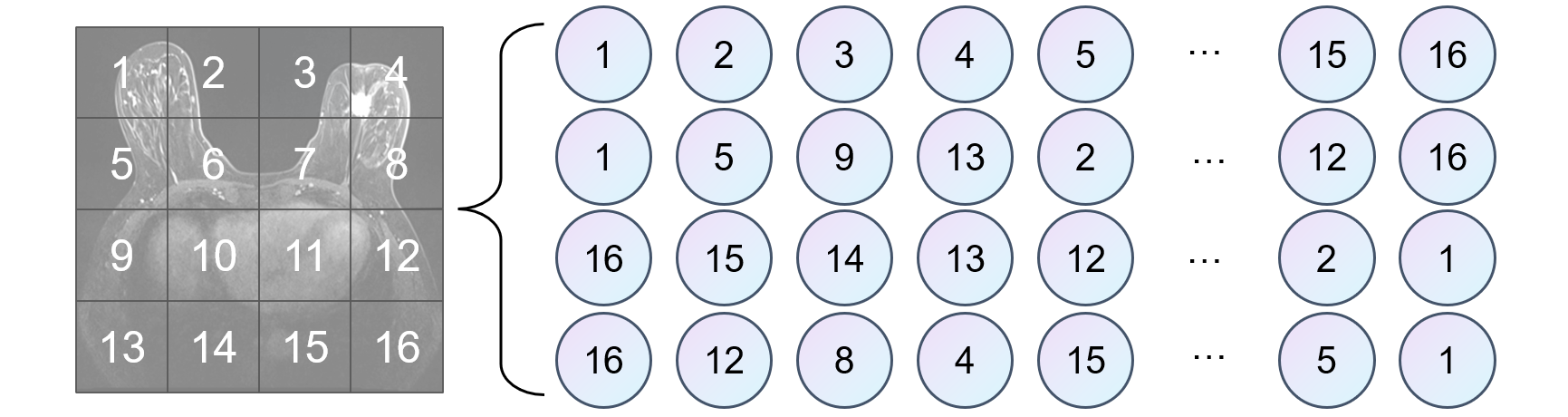}
    \caption{Cross-Scan Mechanism. }
    \vspace{-19pt}
    \label{figure:CSM}
    \end{figure}
\begin{itemize}
\item We propose a Hybrid CNN-Mamba with Axial Self-Attention UNet (HCMA-UNet), which integrates CNN’s local feature perception with Mamba’s superior sequential modeling capability for efficient 3D breast cancer segmentation in DCE-MRI.

\item We design a Feature-guided Region-aware Loss (FRLoss) to address the challenge of high similarity between positive and negative pixels in breast cancer images, improving the model’s ability to distinguish subtle differences and enhance segmentation accuracy.

\item HCMA-UNet achieves state-of-the-art performance with reduced computational complexity on three datasets. Moreover, the proposed FRLoss demonstrates superior segmentation accuracy and robust cross-architecture generalization performance.

\end{itemize}

\section{Methods}

    The proposed HCMA-UNet network architecture is illustrated in Fig.~\ref{fig:HCMA}. It follows an encoder-decoder structure, where the encoder integrates Res Block and Hybrid Block (consisting of MISM and Dense Block). The decoder combines HCMA-Up (single convolution block) with transpose convolutions, and its final output is used to calculate FRLoss with the labels.
    \subsection{Res Block}
    We propose a lightweight residual block that builds upon MobileNetV2~\cite{sandler2018mobilenetv2} and ResNet-D~\cite{he2019bag} for downsampling. We adapted the ResNet-D architecture by replacing the feature extraction path with depthwise convolutions from MobileNetV2 and removing the pointwise convolution in the residual path, resulting in substantial parameter reduction while maintaining model effectiveness.
    \subsection{MISM}
     To enhance Mamba's performance in 3D medical image segmentation, we propose the MISM. Our approach reslices 3D volumes into three orthogonal sets of 2D slices (axial, coronal, and sagittal), enabling comprehensive spatial correlation capture through multi-view analysis. The structure of MISM as illustrated in Fig.~\ref{fig:MISM}. MISM incorporates three synergistic components:(1) VSSB, which performs efficient bidirectional scanning operations along both the height and width dimensions of each 2D slice, effectively capturing the intra-slice features. (2) The ASA mechanism models inter-slice dependencies by learning relationships between slices to preserve critical anatomical correlations across different layers. (3) The ASC strategy is proposed to address the computational redundancy of processing identical feature maps in three directions. Given that the axial plane simultaneously displays the complete structure and symmetrical features, we allocate 50\% of the channels to axial plane learning, while equally distributing the remaining channels (25\% each) to coronal and sagittal planes.

    \subsubsection{VSSB}
     VMamba~\cite{liu2024vmamba} achieves breakthroughs in 2D visual representation learning by introducing the VSSB (combining the SS2D module with the S6 block). The SS2D employs a Cross-Scan Mechanism (CSM) (Fig.~\ref{figure:CSM}), which performs bidirectional scanning along both the horizontal and vertical dimensions, thereby enabling each pixel to aggregate features through forward and backward paths for comprehensive spatial context modeling.
    \subsubsection{ASA}
    To effectively capture the spatial dependencies between slices in medical volumes, we propose ASA, a one-dimensional Self-Attention mechanism. Operating along the direction orthogonal to the VSSB scanning plane, ASA achieves linear computational complexity. The query (Q), key (K), and value (V) features are generated through pointwise convolution from the input tensor $x \in \mathbb{R}^{B \times C \times L_1 \times L_2 \times L_3}$, where $B$ and $C$ denote batch size and number of channels respectively, and $L_1$, $L_2$, $L_3$ are the spatial dimensions. To capture inter-slice relation, we reshape these features along the $L_3$ direction (similar operations can be applied to the $L_1$ or $L_2$ direction) into the following dimensions:
    \begin{equation}
    \begin{aligned}
         Q' &= \text{reshape}\left(Q\right) \in \mathbb{R}^{(B \cdot L_1 \cdot L_2) \times L_3 \times C} \\
         K' &= \text{reshape}\left(K\right) \in \mathbb{R}^{(B \cdot L_1 \cdot L_2) \times C \times L_3} \\
         V' &= \text{reshape}\left(V\right) \in \mathbb{R}^{(B \cdot L_1 \cdot L_2) \times L_3 \times C} \\
    \end{aligned}
    \end{equation}
    \indent Then, we compute the attention result using scaled dot-product attention:
    \begin{equation}
        Result = softmax\left(\frac{Q'K'}{\sqrt{C}}\right)V'   
    \end{equation}
    \vspace{-24pt}
    \subsection{Dense Block}
    Inspired by the memory-efficient dense connectivity design of RDNet~\cite{kim2025densenets}, we propose a densely connected convolution module that enhances feature representations through progressive feature aggregation. Given an input feature $x$, the module first applies a 3×3×3 depthwise convolution to obtain the transformed feature $x_1$. The original feature $x$ and the transformed feature $x_1$ are then concatenated along the channel dimension, followed by a pointwise convolution to generates $x_2$. Finally, $x$, $x_1$, $x_2$ are concatenated and processed by a pointwise convolution to produce the final output with the desired channel dimensions. This dense connectivity pattern enables effective feature reuse while maintaining computational efficiency through the use of depthwise and pointwise convolutions.
    \vspace{-5pt}
    \subsection{FRLoss}
        To address the challenges of feature discrimination in medical image segmentation, we propose a Feature-guided Region-aware Loss (FRLoss) that operates on the feature maps $F\in\mathbb{R}^{B \times C \times D \times H \times W}$ before the Out Block, with its calculation process illustrated in the Loss Calculation region of Fig.~\ref{fig:HCMA} . This loss function guides feature learning through three complementary components: Positive Compactness Loss, Boundary Aware Loss and Hard Negative Pixels Mining Loss.
        \subsubsection{Foreground Center}
        We compute the Foreground Center $f_p$ by averaging the features of all foreground pixels. It is calculated as follows:
        \begin{equation}
        \setlength{\abovedisplayskip}{6pt}
        \setlength{\belowdisplayskip}{6pt}
        f_p = \frac{1}{N_{pos}}\sum_{i=1}^{N_{pos}}f_i
        \end{equation}
        Where $N_{\mathrm{pos}}$ denotes the total number of positive pixels, and $f_i \in \mathbb{R}^C$ represents the feature vector of the i-th foreground pixel extracted from feature maps before Output Block.
        \subsubsection{Positive Compactness Loss}
            To enhance the feature compactness of breast lesions, we propose a Feature-guided Positive Compactness Loss that leverages the mean lesion feature to guide individual lesion features in the feature maps, formulated as:
            \begin{equation}
            \setlength{\abovedisplayskip}{6pt}
            \setlength{\belowdisplayskip}{6pt}
            \mathcal{L}_{\text{pos}} = \frac{1}{N_{\text{pos}}} \sum_{i=1}^{N_{\text{pos}}} \left(1 - sim \left (f_i,f_p \right ) \right)
            \end{equation}
             Where $ sim \left (\cdot \right )$ represents the computation of cosine similarity.   
        \subsubsection{Boundary Aware Loss}
            To enhance boundary segmentation accuracy, we propose a Boundary Aware Loss. The computation process consists of the following key steps: First, we obtain the potential boundary confusion region $\mathcal{M}$ by performing $T_1$ times dilation with a 3×3×3 kernel on positive pixel set and intersecting with negative pixel regions; Then, we calculate the cosine similarity between the features in this region and the Foreground Center, formulated as:
            \begin{equation}
            \begin{aligned}
                \mathcal{M} &= \text{dilate}^{T_{1}}(\Omega_{\text{pos}}) \cap {\Omega_{\text{neg}}}\\
                \mathcal{L}_{\text{boundary}} &= \frac{1}{|\mathcal{M}|} \sum_{i \in \mathcal{M}} \sigma_r \left ( sim \left (f_i,f_p \right ) \right)
            \end{aligned}
            \end{equation}
        Where $\Omega_{\text{neg}}$ and $\Omega_{\text{pos}}$ denote negative and positive pixel sets respectively. $\sigma_r(\cdot)$ is the ReLU function.
        \subsubsection{Hard Negative Pixels Mining Loss}
            We propose a Hard Negative Pixels Mining Loss to address challenging negative regions where breast tissues and other organs often exhibit similar appearances to breast tumors. These confusing regions frequently lead to false positive predictions, thus significantly reducing the model's precision during the segmentation process. By penalizing negative regions with high feature similarity to positive pixels, this loss reduces the likelihood of false positive predictions.
           To reduce the computational cost, we select $N$ negative pixels that are most similar to the positive pixels. We first compute the cosine similarity between each negative pixel and Foreground Center, formulated as
            \begin{equation}
            \mathcal{S} = \{s_i = sim \left (f_i,f_p \right ) \mid i \in \Omega_{\text{neg}}\}
            \end{equation}

            The negative pixels with high similarity scores to positive pixels often correspond to confounding anatomical structures that share similar visual characteristics with the target organ. For the top-$N$ negative pixels with the highest similarity scores, we apply dilation operations $T_2$ times and calculate their intersection with $\mathcal{N}$ to identify these challenging regions, formulated as:
            \begin{equation}
            \mathcal{N} = \text{dilate}^{T_2}({\text{top}_N(\mathcal{S})}) \cap \Omega_{\text{neg}}
            \end{equation}
            
            Based on the obtained negative pixel set $\mathcal{N}$, we construct the loss function by computing the cosine similarity between the features of each pixel in the set and those of the Foreground Center, formulated as:
            \begin{equation}
            \mathcal{L}_{\text{neg}} = \frac{1}{|\mathcal{N}|} \sum_{i \in \mathcal{N}} \sigma_r \left ( sim \left (f_i,f_p \right ) \right)
            \end{equation}
        
    \subsubsection{Overall Loss Function}
    FRLoss is employed in combination with conventional Dice Loss and Cross Entropy Loss, and the overall loss function can be formulated as follows:
    \begin{equation}
        \begin{aligned}
        \mathcal{L}_{\text{total}} &= \mathcal{L}_{\text{Ce}} + \mathcal{L}_{\text{Dice}} + \lambda \mathcal{L}_{\text{FR}} \\
        \mathcal{L}_{\text{FR}} & = \mathcal{L}_{\text{pos}}  + \mathcal{L}_{\text{boundary}} + \mathcal{L}_{\text{neg}}
        \end{aligned}
    \end{equation}

    Where $\lambda$ denotes the weight of FRLoss.
\begin{table*}[!h]
\renewcommand{\arraystretch}{0.85}
\centering
\caption{Quantitative comparison of different methods on breast cancer segmentation tasks, where HCMA-UNet and FRLoss are our proposed network and loss function, respectively. Results are evaluated on one private dataset (Dataset I) and two public datasets (Dataset II and III).}
\vspace{0 pt}

\setlength{\tabcolsep}{0.8mm}{
\begin{tabular}{@{}cc|ccccc|ccccc|ccccc|cc@{}}
\toprule
\multirow{2}{*}{\textbf{Models}} & \multirow{2}{*}{\textbf{Loss}} & \multicolumn{5}{c|}{\textbf{Dataset \uppercase\expandafter{\romannumeral1}
}} & \multicolumn{5}{c|}{\textbf{Dataset \uppercase\expandafter{\romannumeral2}
}} & \multicolumn{5}{c|}{\textbf{Dataset \uppercase\expandafter{\romannumeral3}
}} & \multirow{2}{*}{\textbf{MParams}} & \multirow{2}{*}{\textbf{GFLOPs}} \\
& & \textbf{Dice} & \textbf{Prec} & \textbf{Rec} & \textbf{IoU} & \textbf{VS} & \textbf{Dice} & \textbf{Prec} & \textbf{Rec} & \textbf{IoU} & \textbf{VS} & \textbf{Dice} & \textbf{Prec} & \textbf{Rec} & \textbf{IoU} & \textbf{VS} & & \\

\midrule

\multirow{2}{*}{Attention Unet~\cite{SCHLEMPER2019197}} & w/o FRLoss & 75.42 & 78.54 & 77.33 & 66.45 & 82.37 & 72.20 & 73.90 & 79.86 & 61.13 & 81.57 & 68.93 & 73.92 & 70.88 & 60.27 & 75.73 & \multirow{2}{*}{16.66} & \multirow{2}{*}{1148.28} \\
& w/ FRLoss & 77.46 & 78.99 & 80.27 & 68.51 & 84.95 & 75.36 & \textbf{79.80} & 78.96 & 64.97 & 82.52 & 72.84 & 78.63 & 72.44 & 63.55 & 81.85 & & \\

\midrule

\multirow{2}{*}{nnU-Net~\cite{isensee2021nnu}} & w/o FRLoss & 76.85 & 78.58 & 80.06 & 67.34 & 84.82 & 73.84 & 75.32 & 80.26 & 62.79 & 82.27 & 70.43 & 74.34 & 75.08 & 61.42 & 80.46 & \multirow{2}{*}{30.45} & \multirow{2}{*}{2610.66} \\
& w/ FRLoss & 77.85 & 79.88 & 80.65 & 68.52 & 85.24 & 75.42 & 77.94 & 80.50 & 64.92 & 82.93 & 72.14 & 80.62 & 74.23 & 63.09 & 78.37 &  &  \\

\midrule

\multirow{2}{*}{MedNeXt~\cite{roy2023mednext}}\ & w/o FRLoss & 76.25 & 77.44 & 79.50 & 67.09 & 83.14 & 73.11 & 73.64 & 80.88 & 61.97 & 81.90 & 71.39 & 77.85 & 74.34 & 61.96 & 78.61 & \multirow{2}{*}{17.60} & \multirow{2}{*}{525.70} \\
& w/ FRLoss & 78.22 & 77.83 & 83.02 & 68.95 & 85.45 & 75.20 & 77.54 & 79.93 & 64.90 & 83.12 & 74.19 & 76.51 & 77.59 & 64.35 & 81.79 &  &  \\

\midrule

\multirow{2}{*}{UX-Net~\cite{lee20223d}} & w/o FRLoss & 75.30 & 78.36 & 79.30 & 65.86 & 81.90 & 69.14 & 68.24 & 79.58 & 57.20 & 79.61 & 70.73 & 74.08 & 72.47 & 60.31 & 80.41 & \multirow{2}{*}{34.71} & \multirow{2}{*}{1349.14} \\
& w/ FRLoss & 75.76 & 78.77 & 79.79 & 66.24 & 82.40 & 72.66 & 72.91 & 80.48 & 61.08 & 82.02 & 70.86 & 73.60 & 72.57 & 61.22 & 81.12 &  &  \\

\midrule

\multirow{2}{*}{UNETR++~\cite{shaker2024unetr++}} & w/o FRLoss & 76.15 & 76.98 & 80.51 & 66.67 & 82.36 & 68.82 & 69.25 & 78.49 & 57.30 & 79.15 & 72.72 & 77.48 & 72.90 & 62.75 & 84.88 & \multirow{2}{*}{42.64} & \multirow{2}{*}{272.48} \\
& w/ FRLoss & 76.88 & 79.38 & 80.67 & 67.32 & 83.89 & 69.59 & 70.13 & 79.37 & 58.08 & 78.71 & 73.10 & 78.90 & 72.53 & 63.32 & 81.42 &  &  \\

\midrule

\multirow{2}{*}{SwinUNETR~\cite{hatamizadeh2021swin}} & w/o FRLoss & 75.57 & 77.64 & 78.64 & 66.12 & 83.19 & 64.58 & 62.66 & 79.78 & 52.70 & 74.50 & 68.06 & 68.20 & 79.40 & 57.48 & 75.84 & \multirow{2}{*}{15.64} & \multirow{2}{*}{394.96} \\
& w/ FRLoss & 76.84 & 79.13 & 79.30 & 67.16 & 84.08 & 71.53 & 72.25 & 79.82 & 60.22 & 80.55 & 70.73 & 70.62 & 74.85 & 61.22 & \textbf{86.89} &  &  \\

\midrule

\multirow{2}{*}{LKM-UNet~\cite{wang2024lkm}} & w/o FRLoss & 76.31 & 77.19 & 81.33 & 66.82 & 83.37 & 72.08 & 72.61 & 80.27 & 60.77 & 81.11 & 67.83 & 76.74 & 68.88 & 58.98 & 75.74 & \multirow{2}{*}{102.19} & \multirow{2}{*}{4938.70} \\
& w/ FRLoss & 76.69 & 77.21 & 81.21 & 66.94 & 84.39 & 73.58 & 74.41 & 80.22 & 62.63 & 82.17 & 68.75 & 76.85 & 71.59 & 60.26 & 75.67 & & \\
\midrule

\multirow{2}{*}{UMamba~\cite{ma2401u}} & w/o FRLoss & 76.67 & 76.97 & 81.39 & 67.26 & 84.91 & 69.83 & 69.77 & 80.74 & 58.18 & 79.14 & 70.07 & 71.55 & 74.15 & 61.67 & 79.17 & \multirow{2}{*}{69.37} & \multirow{2}{*}{13447.52} \\
& w/ FRLoss & 77.83 & 79.02 & 81.46 & 68.15 & 84.56 & 75.36 & 77.28 & 80.27 & 64.90 & 82.77 & 74.80 & 79.40 & 76.90 & 64.50 & 82.50 & & \\
\midrule

\multirow{2}{*}{MambaClinix~\cite{bian2024mambaclinix}} & w/o FRLoss & 78.01 & 78.78 & 83.08 & 68.09 & 85.32 & 69.06 & 68.20 & 81.75 & 57.54 & 77.68 & 72.06 & 75.07 & 75.55 & 62.85 & 80.30 & \multirow{2}{*}{108.33} & \multirow{2}{*}{13960.54} \\
& w/ FRLoss & 78.43 & 78.93 & 83.75 & 68.95 & 85.16 & 69.70 & 68.03 & \textbf{81.91} & 58.22 & 78.65 & 75.53 & 80.83 & \textbf{77.32} & 65.76 & 75.53 & & \\
\midrule

\multirow{2}{*}{HCMA-UNet(Ours)} & w/o FRLoss & 79.27 & 79.20 & 84.70 & 69.04 & 86.54 & 74.41  & 75.64 & 80.43 & 63.23 & 83.43 & 77.92 & 80.34 & 76.61 & 68.15 & 86.75 & \multirow{2}{*}{\textbf{2.87}} & \multirow{2}{*}{\textbf{126.44}} \\
& w/ FRLoss & \textbf{80.28} & \textbf{80.23} & \textbf{85.37} & \textbf{70.16} & \textbf{87.49} & \textbf{76.14} & 77.84 & 81.39 & \textbf{65.46} & \textbf{84.04} & \textbf{78.05} & \textbf{80.87} & 76.65 & \textbf{68.52} & 86.25 \\
\bottomrule
\label{tab:Compare}
\end{tabular}}
\vspace{-15 pt}
\end{table*}

\section{Experiments and Results}
\subsection{Private Dataset}
Our main research dataset (Dataset \uppercase\expandafter{\romannumeral1}) consists of 907 patient imaging data collected from two medical centers. The lesion segmentation annotations are validated by experienced physicians through a double-checking process. We randomly split the dataset into training and test sets with 725 and 182 samples respectively.
\subsection{Public Dataset}
MAMA-MIA (Dataset \uppercase\expandafter{\romannumeral2})~\cite{garrucho2024mama} dataset contains 1506 breast cancer DCE-MRI cases with expert-annotated tumor segmentations. We use the first post-contrast images and split them into 1204 training and 302 test samples.

The second external dataset (Dataset \uppercase\expandafter{\romannumeral3}) is a publicly available DCE-MRI dataset~\cite{zhang2023robust} from Yunnan Cancer Hospital. The post-contrast enhanced images are selected and randomly partitioned into 75 training and 25 test samples.
\subsection{Implementation Details}
 All experiments are conducted using PyTorch on an NVIDIA RTX 3090 GPU. Data preprocessing and augmentation follow the nnUNet framework. During training, we use patches of $128^3$ with batch size 2. Two loss configurations are employed: (1) a weighted combination of Dice and CE loss; (2) the proposed FRLoss ($T_1 = 10$, $T_2 = 10$, $N = 250$, $\lambda = 5$) added to configuration 1. The network is trained 
 for 500 epochs using AdamW optimizer with a learning rate of 1e-4. Performance is evaluated using Dice, IoU, Precision (Prec), Recall (Rec), and Volumetric Similarity (VS) metrics, along with model efficiency metrics including the number of parameters (in millions) and computational complexity (in GFLOPs).
             \begin{figure}[!htbp]
            \centering
            \includegraphics[width=0.48\textwidth]{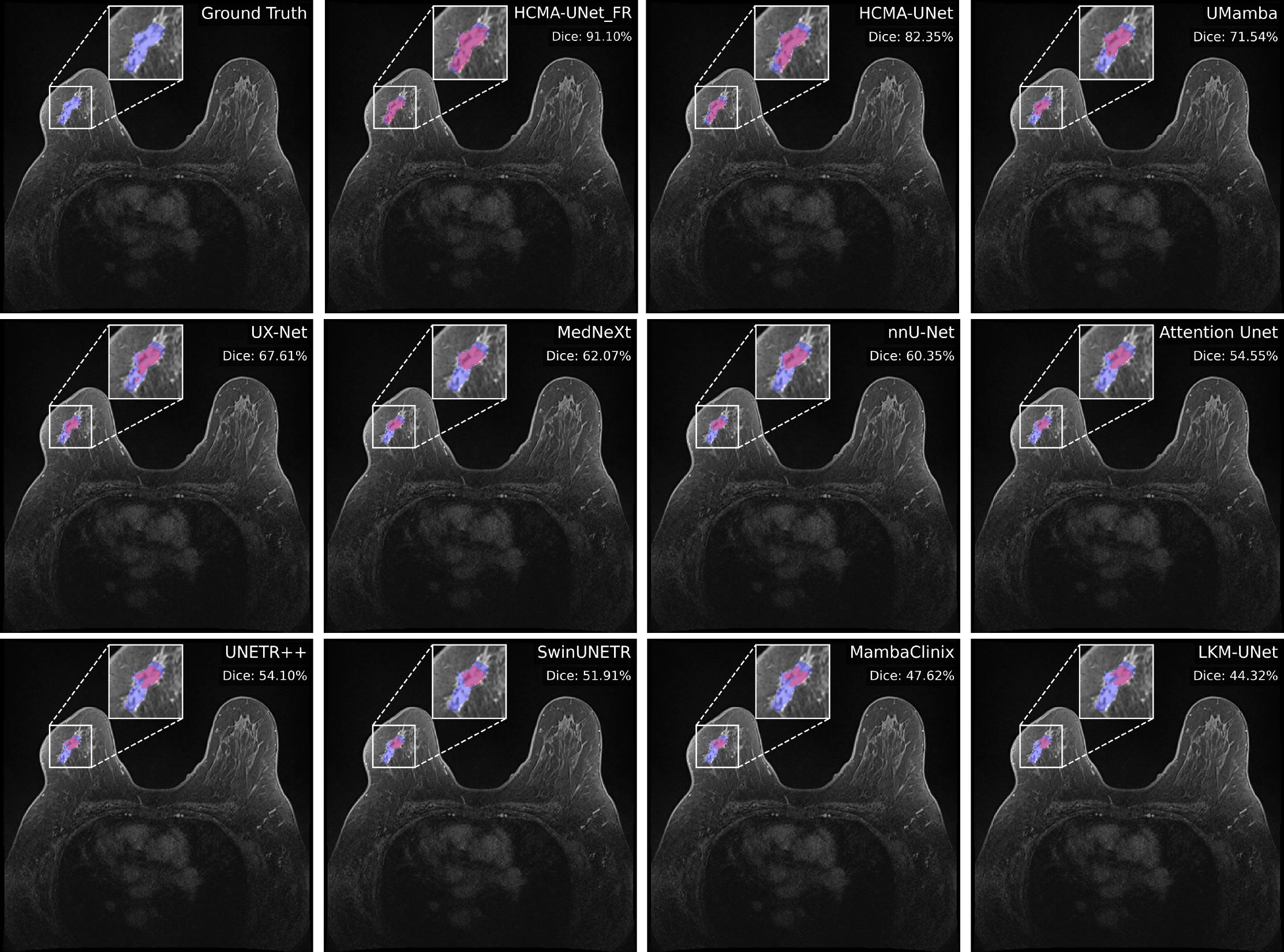}
            \caption{Visual comparison of segmentation results across ten different methods on a representative axial slice from our private dataset. Models marked with FR indicate training with FRLoss.}
            \vspace{-14pt}
            \label{fig:SegResult}
            \end{figure}
 \vspace{-4pt}
 \subsection{Result Comparison} 
As shown in Table~\ref{tab:Compare}, we comprehensively evaluate our proposed HCMA-UNet against nine mainstream medical image segmentation models, using a hybrid Dice and CE Loss as the baseline. The experimental results demonstrate that HCMA-UNet achieves superior performance with significant improvements in multiple evaluation metrics, reaching Dice scores of 80.28\%, 76.14\%, and 78.05\% on Datasets I, II, and III respectively. The proposed FRLoss consistently enhances the performance of various network architectures, demonstrating its strong generalization capability. Furthermore, HCMA-UNet shows remarkable computational efficiency with only 2.87M parameters and 126.44 GFLOPs, which is substantially lower than UMamba (69.37M parameters, 13447.52 GFLOPs) and nnU-Net (30.45M parameters, 2610.66 GFLOPs), significantly reducing the computational resources required for medical image segmentation while maintaining high segmentation quality. 

Fig.~\ref{fig:SegResult} illustrates the segmentation results on a challenging case with ambiguous boundaries. On this challenging slice, our proposed HCMA-UNet achieves a Dice coefficient of 82.35\%, demonstrating a substantial improvement of 10.81\% over the best-performing existing method, UMamba. Furthermore, by incorporating FRLoss as the loss function, HCMA-UNet further enhances its segmentation performance to 91.10\%, accurately delineating the complete lesion region.
\begin{table}[!h]
\vspace{-10pt}
\renewcommand{\arraystretch}{1.0} 
\caption{Ablation Study of MISM on Dataset \uppercase\expandafter{\romannumeral1}}
\setlength{\tabcolsep}{0.76mm}{
\begin{tabular}{@{}cccc|ccccc|c@{}}
\toprule
\textbf{VSSB} & \textbf{ASC} & \textbf{ASA} & \textbf{Mamba3d} & \textbf{Dice} & \textbf{Prec} & \textbf{Rec} & \textbf{IoU} & \textbf{VS} & \textbf{MParams}\\
\midrule
 - & - & - & - & 77.87 & 79.10 & 81.11 & 68.24 & 85.66 & 2.58  \\
 \checkmark & - & - & - & 78.52 & 78.58 & 84.68 & 68.67 & 85.75 & 3.19  \\
 \checkmark & \checkmark & - & - & 78.66 & 79.12 & 83.02 & 68.50 & 86.31 & 2.76  \\
 \checkmark & - & \checkmark & - & 79.16 & 78.86 & \textbf{84.90} & \textbf{69.34} & 86.40 & 4.01  \\
 \checkmark & \checkmark & \checkmark & - & \textbf{79.27} & \textbf{79.20} & 84.70 & 69.04 & \textbf{86.54} & 2.87  \\
 - & - & - & \checkmark & 78.09 & 78.19 & 82.07 & 68.09 & 86.02 & 3.22 \\
\bottomrule
\vspace{-15pt}
\end{tabular}}
\label{tab:MISM_Aba}
\end{table}
\subsection{Ablation experiment of MISM}
 To evaluate the performance of the HCMA-UNet network and the effectiveness of MISM comprehensively, we design six ablation experiments. The experiments primarily focus on three core components of MISM: VSSB, ASC strategy, and ASA. The experimental results are listed in Table~\ref{tab:MISM_Aba}.  First, we construct a baseline network containing only convolutional layers. The experimental results demonstrate that as various components are gradually introduced, the network's performance metrics exhibit a continuous upward trend. Among them, the complete HCMA-UNet architecture exhibits optimal performance. To compare with the direct application of bidirectional Mamba, we replace MISM with a 3DMamba block. Our method outperforms 3DMamba with improvements of 1.18\%, 1.01\%, and 2.63\% in Dice, Precision, and Recall, respectively. These improvements demonstrate MISM's effectiveness in medical images.
 \begin{table}[H]
 \vspace{-9pt}
\renewcommand{\arraystretch}{1.0}
\centering
\caption{Ablation Study of Loss Function on Dataset \uppercase\expandafter{\romannumeral1}}
\setlength{\tabcolsep}{0.65mm}{
\begin{tabular}{@{}lccc|ccccc@{}}
\toprule
  \textbf{Loss} & $\mathcal{L}_{\text{pos}}$ &  $\mathcal{L}_{\text{boundary}}$ & $\mathcal{L}_{\text{neg}}$ & \textbf{Dice} & \textbf{Prec} & \textbf{Rec} & \textbf{IoU} & \textbf{VS} \\
\midrule
 Dice+CE &- & - &  -  & 79.27 & 79.20 & 84.70 & 69.04 & 86.54 \\
    $\mathcal{L}_{\text{pos}}$ & \checkmark & - & - & 79.58 & 79.49 & 84.73 & 69.67 & 86.67 \\
    $\mathcal{L}_{\text{boundary}}$ & - & \checkmark & - & 79.74 & 79.61 & 84.78 & 69.64 & 87.12 \\
    $\mathcal{L}_{\text{neg}}$ & - & - & \checkmark & 79.64 & 80.07 & 84.61 & 69.48 & 86.31 \\
    $\mathcal{L}_{\text{pos}}$+$\mathcal{L}_{\text{boundary}}$ & \checkmark & \checkmark & - & 79.96 & 79.89 & 85.15 & 69.75 & 86.99 \\
    $\mathcal{L}_{\text{pos}}$+$\mathcal{L}_{\text{neg}}$ & \checkmark & - & \checkmark & 80.09 & 80.05 & 84.98 & 70.09 & 87.08 \\
    $\mathcal{L}_{\text{boundary}}$+$\mathcal{L}_{\text{neg}}$ & - & \checkmark & \checkmark & 79.86 & 79.73 & 84.83 & 69.76 & 87.31 \\
$\mathcal{L}_{\text{pos}}$+$\mathcal{L}_{\text{boundary}}$+$\mathcal{L}_{\text{neg}}$ & \checkmark & \checkmark & \checkmark & \textbf{80.28} & \textbf{80.23} & \textbf{85.37} & \textbf{70.16} & \textbf{87.49} \\
\bottomrule
\end{tabular}}
\label{tab:Loss_Aba}
\end{table}
 \subsection{Ablation experiment of FRLoss} To evaluate the contribution of each component in FRLoss, we conduct ablation experiments on our three proposed loss terms. As presented in Table~\ref{tab:Loss_Aba}, the Positive Compactness Loss improves IoU by 0.63\%, validating its effectiveness in enhancing the spatial cohesion of positive pixels. The Boundary Aware Loss achieves a Dice of 79.74\%, demonstrating its capability in precise boundary delineation. The Hard Negative Pixels Mining Loss elevates the Precision to 80.07\%, significantly improving the model's ability to distinguish high-similarity negative pixels in complex regions. When integrating all components, the complete FRLoss architecture demonstrates substantial improvements over the baseline model, with gains of 1.01\%, 1.03\%, 0.67\%, and 1.12\% in Dice, Precision, Recall, and IoU metrics, respectively, confirming the complementary effectiveness of our proposed loss design.

\section{Conclusion}
In this work, we propose HCMA-UNet for breast cancer lesion segmentation in DCE-MRI, which integrates our MISM module with a lightweight CNN backbone for efficient local and long-range feature modeling. We also introduce FRLoss, a composite loss function designed to enhance feature consistency, boundary delineation, and reduce false positives. Experiments demonstrate that HCMA-UNet achieves state-of-the-art performance across multiple datasets while maintaining lower computational costs. Extensive experiments of FRLoss on various network architectures and datasets demonstrate its strong generalization ability.
\bibliographystyle{IEEEbib}
\bibliography{ref}

\end{document}